# Phononic band structure engineering for high-*Q* gigahertz surface acoustic wave resonators on lithium niobate

*Linbo Shao[1]†\*, Smarak Maity[1]†, Lu Zheng[2], Lue Wu[1], Amirhassan Shams-Ansari[1], Young-Ik Sohn[1], Eric Puma[1], M. N. Gadalla[1], Mian Zhang[1], Cheng Wang[1], Keji Lai[2], Marko Lončar[1]\**

[1]*John A. Paulson School of Engineering and Applied Sciences, Harvard University, 29 Oxford Street, Cambridge, MA 02138, USA*
[2]*Department of Physics, University of Texas at Austin, Austin, Texas 78712, USA*

*† These authors contributed equally to this work.*
*\* Email: shaolb@seas.harvard.edu (L.S.); loncar@seas.harvard.edu (M.L.)*

*Abstract*
*Phonons at gigahertz frequencies interact with electrons, photons, and atomic systems in solids, and therefore have extensive applications in signal processing, sensing, and quantum technologies. Surface acoustic wave (SAW) resonators that confine surface phonons can play a crucial role in such integrated phononic systems due to small mode size, low dissipation, and efficient electrical transduction. To date, it has been challenging to achieve high quality (Q) factor and small phonon mode size for SAW resonators at gigahertz frequencies. Here, we present a methodology to design compact high-Q SAW resonators on lithium niobate operating at gigahertz frequencies. We experimentally verify out designs and demonstrate Q factors in excess of $2\times10^4$ at room temperature ($6\times10^4$ at 4 Kelvin) and mode area as low as 1.87 $\lambda^2$. This is achieved by phononic band structure engineering, which provides high confinement with low mechanical loss. The frequency-Q products (fQ) of our SAW resonators are greater than $10^{13}$. These high-fQ and small mode size SAW resonators could enable applications in quantum phononics and integrated hybrid systems with phonons, photons, and solid-state qubits.*

**Introduction**

Microwave phonons have attracted interest in quantum technologies as an efficient and versatile means of coupling superconducting qubits[1-7], solid-state quantum defects[8-13], microwave fields[14-18], and optical photons[19-23], and thus have emerged as a promising platform for the realization of quantum networks. Surface acoustic waves (SAW) are acoustic waves propagating on the surface of an elastic solid with amplitude decaying into the solid [24]. In the view of a phononic band structure, SAW are phonon modes confined to solid surfaces due to the phase mismatch with bulk modes. SAW can couple efficiently to electromagnetic fields via the piezoelectric effect, and has been used to filter signals for 5G communications[25], modulate light via the acousto-optic effect[19-21], and drive solid-state electronic spins through spin-orbit coupling[10,11]. Compared to other micromechanical resonators, including suspended optomechanical nanobeam cavities[22,26-31], bulk acoustic wave cavities[1-3], and 2D material micromechanical cavities[32-34], SAW resonators could have one or more advantages of planar fabrication, strong electrical coupling efficiency, high confinement, and gigahertz resonant frequency. Gigahertz Fabry-Perot (FP) acoustic resonators have been demonstrated using phononic crystals[35-38] or distributed Bragg reflectors[15,39-42], but they are not optimized for emerging quantum applications, where higher Q factors and smaller mode sizes are desired to sufficiently enhance the interactions.

Electrically-coupled integrated acoustic devices make use of piezoelectric materials such as quartz[14,15,37], zinc oxide[10,35,36,40], gallium arsenide[6,16,20], gallium nitride[39,41], aluminum nitride[1,11,21,42], and lithium niobate[4,25,38,43,44]. Among these materials, lithium niobate possesses an electromechanical coupling coefficient that is much greater than that of the other materials. At room temperature, the acoustic propagation loss due to electrical conductivity is significantly lower for lithium niobate than for semiconductor materials. Lithium niobate SAW devices[24,43] are used in electrical signal processing (e.g. filters), acousto-optics, and sensing, due to their smaller footprint, better performance, and easier fabrication compared to that of their electronic counterparts. In addition, the nonlinear effects of lithium niobate enable the realization of state-of-the-art optomechanical devices[29,31,45] and efficient electro-optic devices[46-49], demonstrating its unique potential for integrated hybrid systems.

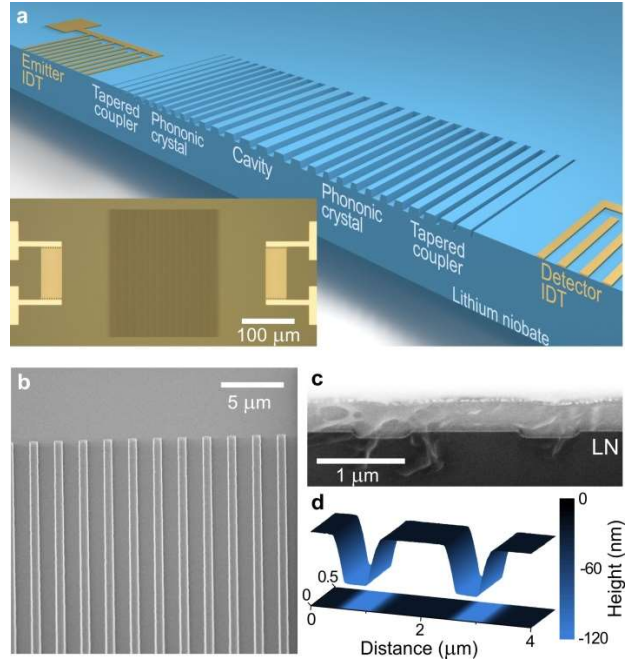

**Figure 1 | Surface acoustic wave resonator on lithium niobate**. **a**, Illustration of band structure engineered surface acoustic resonator on lithium niobate. Inset: optical microscope image of a fabricated device. The dark region at the center is the etched grooves, and the bright regions on the sides are metal IDTs. **b**, Scanning electron microscope image of the cavity part on lithium niobate. **c**, Cross-section showing the etched grooves at the cavity part. A platinum layer is deposited to protect the surface of lithium niobate (LN) from damage during focused ion beam milling. **d**, Surface profile of the SAW resonator scanned by atomic force microscopy.

Here, we experimentally demonstrate high-$Q$ and small mode size SAW resonators at gigahertz frequencies on lithium niobate by engineering phononic band structures using adiabatically-tapered structures. Inspired by the approaches used to design 1D photonic crystal cavities[50], we realize SAW resonators by chirping the period of a quasi-1D phononic crystal (PnC), resulting in $Q$ factors of 16,700 in atmospheric conditions and 61,100 at 4 K. Our resonators also have a small mode area $A_{\text{eff}}$ = 1.87 $\lambda^2$, where $\lambda$ is the wavelength of the SAW on a free surface of lithium niobate. These large $Q$ factors and small mode areas enhance acoustic fields by over three orders of magnitude. The frequency-quality factor products ($fQ$), which indicate the isolation from thermal phonons, reach $2\times10^{13}$ in atmosphere with resonant frequencies ranging from 0.5 to 5 GHz, which is higher than the $fQ$ of the recently reported suspended lithium niobate mechanical resonators[29,44]. The achieved $fQ$ product can allow room-temperature quantum phononics[27], where the minimum required $fQ$ is $6\times10^{12}$ in order to maintain coherence from thermal phonon bath over one mechanical period[51].

**Device principle and fabrication**

We fabricate surface phononic crystal resonators along the X crystal direction on 128°Y-cut lithium niobate [52]. The SAW propagating in this direction experiences low diffraction and efficient coupling to electric fields through piezoelectric effects. Both congruent and black lithium niobate wafers are experimentally tested in this work, and we find no observable difference in acoustic $Q$ factors between them. We choose to use black lithium niobate for its reduced charging effect during electron beam lithography.

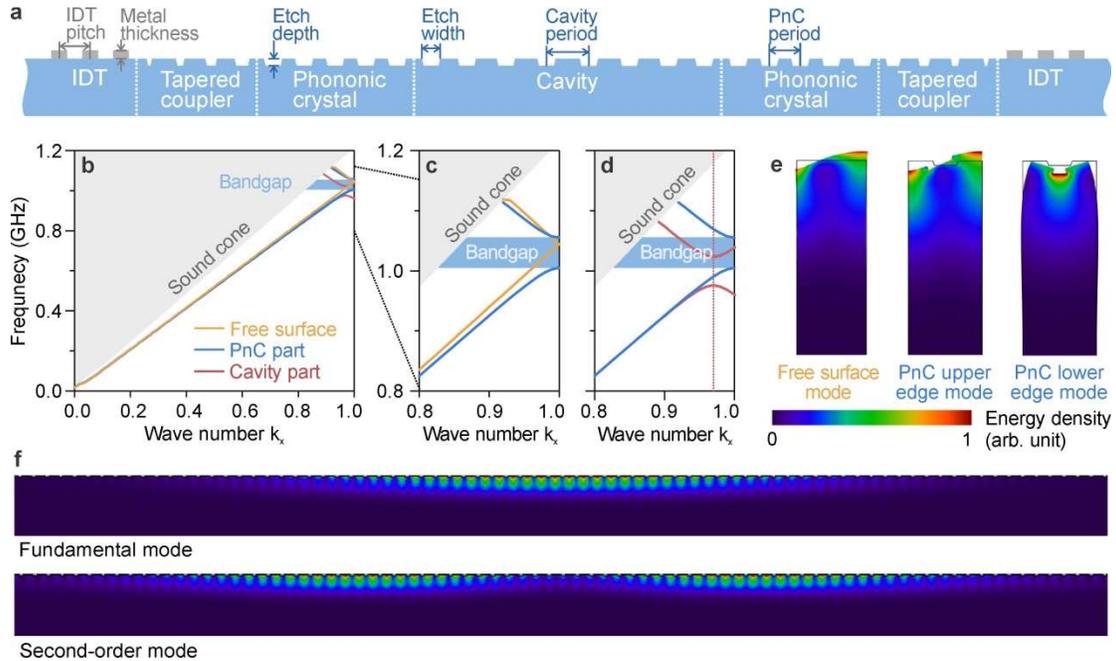

**Figure 2 | Phononic band structure engineering of surface acoustic wave (SAW) on lithium niobate. a**, Schematic of the SAW resonator and its design parameters. **b-d**, Phononic band structures for SAW on unperturbed free surface, surface phononic crystal (PnC) part, and cavity part. The wave number $k_x$ is normalized to the π phase of PnC. The vertical red dash line indicates the π of the cavity part, which is of a greater period. The design parameters are: etch width 0.653 μm, etch depth 115 nm, period for PnC part 1.92 μm, and period for cavity part 1.98 μm. **e**, SAW modes of free surface and PnC with π phase of the unit cells. **f**, Mode profiles of the fundamental and second-order mode of the SAW resonator. The color scale indicates the total energy density of electromagnetic, kinetic, and elastic energy densities.

We engineer the phononic band structures[28,35,36,41,53] and create high-$Q$ resonators with small mode areas by etching grooves into the lithium niobate surface (Fig. 1**a**). These grooves give rise to a band gap in the phononic band structure that can confine the phononic modes. We taper the period of the grooves towards the center to create a SAW resonator. Compared to conventional FP resonators that employ unperturbed free surfaces at the center and Bragg mirrors on the sides, the tapered grooves adiabatically change the reflectivity, resulting in a significantly reduced scattering loss of acoustic waves into the bulk and better confinement of phonons. We also taper the widths of grooves towards the two interdigital transducers (IDT) used to excite the resonator to improve the resonator coupling[28,39]. An optical microscope image of a fabricated device is shown in Fig. 1**a** Inset. Scanning electron microscopy (Fig. 1**b**) confirms that the dimensions of the etched structures agree with design. In the cross sections obtained by focused ion beam milling (Fig. 1**c**), clean and smooth etched surfaces are observed. The surface profiles of the etched structures are measured by atomic force microscopy (Fig. 1**d**).

**Phononic band structure engineering**

We engineer the band structure of SAW by controlling the period, width, and depth of the grooves (Fig. 2**a**). The target frequency for our baseline SAW resonator design is 1 GHz, which corresponds to a SAW wavelength λ of 3.9 µm on a free surface. A period of 1.92 µm, an etch width of 0.65 µm, and an etch depth of 115 nm have been chosen to form a bandgap of ~50 MHz at above 1 GHz in the phononic band structure; the central cavity part has a greater period of 1.98 µm to support the resonant modes (Figs. 2**b** and 2**d**). Generally, a greater etch depth leads to larger bandgap and provides better confinement (Supplementary Fig. 1), but it also increases the scattering loss into the bulk. For many applications[10,11,19], the quality factor divided by mode area Q/A$_{eff}$ is an important figure of merit; it characterizes the buildup factor of the acoustic field in the cavity. The chosen etch depth of 115 nm (2.8% λ) is a good trade-off between a small mode area and a high *Q* factor[8,9].

The band structure of the SAW is relatively insensitive to the etch width (Supplementary Fig. 2), and it is thus challenging to center the resonant mode in the middle of the bandgap by varying the etch width. Therefore, to create the SAW resonator, we vary the period of the grooves but keep the etch width constant. At the center of our resonator, the period of the grooves is increased to align the upper band edge mode to the center of the PnC bandgap (Fig. 2**d**), providing optimum confinement of the resonant mode. We preferably use the upper band edge mode for robustness in fabrication, as it is less sensitive than the lower band edge mode (Supplementary Fig. 1). Further, to achieve a high-*Q* SAW resonator, we adiabatically taper the period of the grooves quadratically over 30 periods from the center, reducing the scattering loss by the PnC structures[50]. In contrast, a free surface as the central cavity part, as in conventional FP-like SAW resonators, would not be ideal for a small mode area, since SAWs on an unperturbed free surface are too close to the upper band edge of the PnC at the X-point, $k_x$=1 (Fig. 2**c**). This is understood by comparing the mode on the free surface to the modes on the band edge of the surface PnC (Fig. 2**e**); the free surface mode and the upper band edge mode of the PnC have primarily up-and-down motion, while the lower band edge mode has a left-and-right motion.

The simulation shows two high-*Q* SAW modes: a fundamental mode at 1.032 GHz and a second-order mode at 1.045 GHz. (Fig. 2**f**). The period of the central cavity part is fine-tuned to align the resonant frequency of the fundamental mode to the center of PnC bandgap. The mode area for the fundamental mode is 28.2 µm$^2$, equal to 1.87 λ$^2$; the mode area for the second-order mode is 34.1 µm$^2$, equal to 2.31 λ$^2$, where λ=3.9 µm is the SAW wavelength on a free lithium niobate surface. These mode areas are significantly smaller than those of conventional FP-like SAW resonators, which are usually at least one order of magnitude greater. The mode area is defined by the surface integral of normalized total energy densities, i.e. $A_{eff} = \frac{\int (E_{em}+E_k+E_{els}) \, dr}{\max\{E_{em}+E_k+E_{els}\}}$, where $E_{em}$ ($E_k$, $E_{els}$) is the electromagnetic (kinetic, elastic) energy density [52]. This definition is consistent with the convention for optical nanocavities. A large spectral distance of 13 MHz between the fundamental and secondary modes is consistent with the small mode area. Simulated *Q* factors of over 10$^7$ of both fundamental and second-order SAW modes are obtained by eigenmode solutions, where only the acoustic loss to the bulk is included by adding perfectly matched layers at the boundaries in simulation. The high *Q* factors indicate that the PnC provides good confinement of SAW and minimizes the leakage into bulk waves. It is important to note that this simulation overestimates the *Q* since it does not take into account fabrication imperfections, crystal defects, scattering from thermal phonons, and finite electrical conductivities[54].

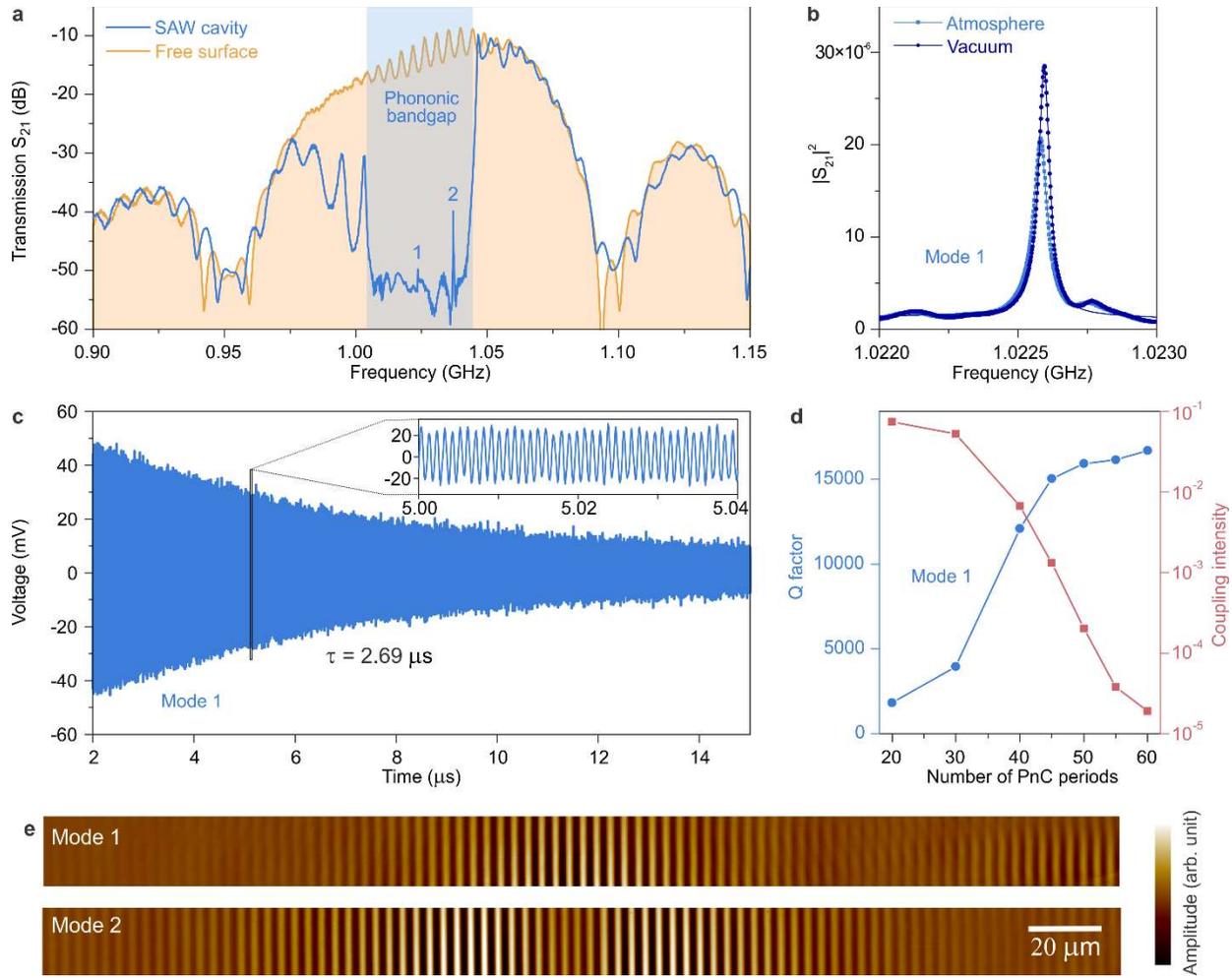

**Figure 3 | Measurements of high-*Q* SAW resonator on lithium niobate. a**, Transmission measurements of a SAW resonator and an unperturbed free surface using a pair of IDT. Two high-*Q* modes, labeled 1 and 2, are observed in the bandgap of the PnC. **b**, Fine scans of transmission measurements for the Mode 1 in atmosphere and vacuum. The measured data (dots) are fitted to Lorentzian peaks (lines). **c**, Resonator ring down measurement of Mode 1 showing a lifetime of 2.69 μs. Inset: the voltage oscillation around 5 μs. **d**, Measurements of Q factors and coupling intensities of Mode 1 of resonators with different numbers of PnC periods. Coupling intensity is measured as the peak height in transmission. **e**, Measurements of SAW mode profiles using transmission-mode microwave impedance microscopy, showing mode profiles as a fundamental mode for Mode 1, and a second-order mode for Mode 2. This is the top view of the device.

**Experimental measurements**

We experimentally characterize the fabricated SAW devices on lithium niobate by measuring the electrical transmission using a vector network analyzer [52]. With periods and metal thicknesses optimized to cover a spectral range of 0.5 to 5 GHz, IDTs are used to excite SAWs and detect their transmission spectra. The IDT bandwidth spans more than 10% of the operating frequency, which allows broadband characterization of the devices. A typical IDT transmission with a bandwidth of 130 MHz is observed for SAW propagating on a free surface (yellow curve in Fig. 3**a**). The small oscillations shown in the transmission are due to the weak resonances induced by the reflections of the metal IDT fingers. For a device with a SAW resonator, the bandgap of the PnC is observed from 1.004 to 1.046 GHz with 40 dB suppression (Blue curve in Fig. 3**a**). The observed bandgap of 41 MHz agrees with the simulation. We speculate that the remaining transmission below -50 dB is due to direct electrical crosstalk between the metal patterns and the probes as well as acoustic leakage though bulk waves. Two high-$Q$ modes are observed in the bandgap labeled mode 1 and 2 in Fig. 3**a**. Mode 1 shows a $Q$ factor of 16,700 at 1.02258 GHz in atmosphere and a $Q$ factor of 21,200 in vacuum (Fig. 3**b**). A ring down measurement in atmosphere gives a lifetime of τ = 2.69 μs for Mode 1, defined as the time taken for the energy to decay to 1/$e$ (Fig. 3**c**). The corresponding $Q$ factor as $Q=\omega_0\tau$=17,300 is in good agreement with the spectral measurements. We experimentally map the SAW mode profiles of devices with the same design using transmission-mode microwave impedance microscopy [55,56]. (Fig. 3**e**) The resonant frequency and the measured mode profile of Mode 1 agree with the fundamental mode shown in Fig. 2**f**. Mode 2 shows a close $Q$ factor of 16,660 at 1.03605 GHz in atmosphere and a $Q$ factor of 20,600 in vacuum (Supplementary Figs. 3**a** and 3**c**). The resonant frequency and the measured mode profile of Mode 2 corresponds to the second-order mode shown in Fig. 2**f**. The observed spectral gap of 13.5 MHz between the fundamental and second-order modes agrees with the numerical simulation. The second-order mode possesses a lower $Q$ factor than the fundamental mode due to less confinement, though it has a higher coupling efficiency. The comparison between $Q$ factors in atmosphere and vacuum indicates air damping to be a significant source of loss.

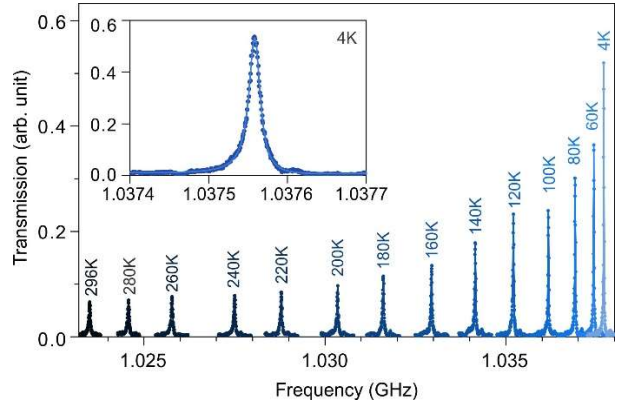

**Figure 4 | Low-temperature measurements of SAW resonators.** Transmission ($|S_{21}|^2$) of the fundamental mode measured during the sample cooling down by continuous flow liquid helium, and the temperature is monitored by a thermal coupler on the sample mount. Inset: A fine scan of the fundamental mode at the stabilized temperature of 4K.

The number of PnC periods surrounding the cavity part determines the loaded $Q$ factors and the coupling strength. Generally, more PnC periods generate better confinement of SAW, leading to higher $Q$ factors, but also reduce the coupling. We fabricate multiple devices with various numbers of PnC periods on a single chip to characterize the relation between $Q$ factors and coupling intensities. For the fundamental mode (Mode 1), the $Q$ factors increase significantly with increasing number of PnC periods, saturating at around 15,000 when more than 45 periods are used (Fig. 3**d**). This saturation indicates that the leakage of the SAW through the PnC is no longer the limit of the $Q$ factors. For the second-order mode (Mode 2), the $Q$ factors improve linearly with increasing number of PnC periods (Supplementary Fig. 3**b**). The coupling intensity reduces exponentially with increasing number of PnC periods, as the mechanical vibration decays exponentially while propagating in the PnC structures.

One key application of our SAW devices is in preparing, controlling, and reading out qubits such as quantum dots, superconducting qubits, and spins. For these applications, functionality at cryogenic temperatures is essential. For this reason, we characterize the low-temperature performance of our SAW resonators using a continuous-flow liquid helium cryostat. The $Q$ factor of the fundamental mode improves from 21,200 at room temperature (Fig. 3**b**) to 61,100 at 4 K (Fig. 4 Inset). Meanwhile, the $Q$ factor for the second-order mode is 42,700 at 4 K, lower than the fundamental mode due to weaker confinement. Since the resonant frequency of 1 GHz corresponds to a temperature of about 50 mK, our devices are in the high temperature ($\hbar\omega \ll k_B T$) regime, where the scattering of acoustic phonons from thermal phonons limits the *Q* factor[54,57], and thus could show even higher *Q* factors at millikelvin temperatures.

The frequency and linewidth of the SAW resonator are monitored continuously during the cooling from room temperature to liquid helium temperature (Fig. 4). The resonant frequency is 1.02350 GHz at room temperature (296 K). It shifts towards higher frequencies by 14 MHz to 1.03743 GHz at 60 K and further by 0.1 MHz to 1.03755 GHz at 4K. On one hand, the thermal expansion coefficient for lithium niobate is temperature dependent[58], and has the value of $14.8\times10^{-6}$/K at 300 K and $1.2\times10^{-6}$/K at 60 K. As the temperature goes down from 300 K to 60 K, the device shrinks about 0.19% geometrically. This shrinkage accounts for a resonant frequency shift of about 2 MHz. On the other hand, the normalized temperature coefficient of the elastic modulus is about $-1.5\times10^{-4}$/K at 298 K (Ref. [59]). Assuming the temperature coefficient decays linearly, the elastic modulus increases by about 1.8% from 300 K to 60 K and could contribute to the 18 MHz shift in the resonant frequency. Thus, the shift towards higher frequencies during the cooling process is primarily caused by the temperature dependence of the elastic modulus.

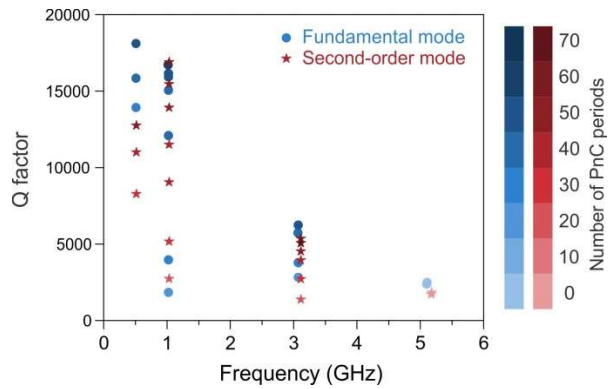

**Figure 5 | *Q* factors on resonant frequencies of SAW resonators in atmosphere**. The fundamental modes and the second-order modes are marked in blue circles and red stars, respectively. The color scale represents the number of PnC periods fabricated to confine the SAW modes.

**SAW resonator scaling**

Our design for SAW resonators can be easily extended to different frequencies by uniform scaling. When all geometric parameters are scaled up by a certain factor, the resonant frequency of the SAW resonator reduces by the same factor. By scaling our 1 GHz design, we demonstrate SAW resonators with resonant frequencies varying over one order of magnitude from 0.5 to 5 GHz (Fig. 5). While the *Q* factors are higher for lower frequency devices, the *fQ* products are similar across the entire range of frequencies. In atmosphere, we observe *Q* factors of 18100, 16700, 6240, and 2480 at fundamental mode frequencies of 0.512, 1.02, 3.07 and 5.11 GHz for scaled SAW devices; and the highest *fQ* product of $2\times10^{13}$ is observed for the device with the median frequency of 3.07 GHz. Due to weaker confinement, the *fQ* products for the second-order modes are lower than the fundamental modes with the same number of PnC periods. For high-frequency devices, though we can scale the geometric design easily, the roughness due to nanofabrication does not scale, and thus roughness induced scattering loss is more significant for higher frequency SAW resonators. In atmosphere, the air damping clamps the *Q* factors at tens of thousands and reduces the *fQ* product for lower frequency devices, e.g., below 0.5 GHz (Ref.[60]).

**Conclusion and outlook**

We demonstrate a systematic method for the design and fabrication of high-$Q$ SAW resonators with small mode areas on lithium niobate. Our phononic band structure engineering optimizes the $Q/A_{eff}$ of SAW resonators by maximizing the confinement of SAW while minimizing the scattering loss into the bulk. These high $Q/A_{eff}$ SAW resonators enhance the strong interaction between phonons and other fields. The demonstrated resonators with $fQ$ products over $10^{13}$ at resonant frequencies ranging from 0.5 to 5 GHz allow potential room-temperature quantum phononics on the surface of lithium niobate. Furthermore, our SAW resonators have mode areas down to 1.87 $\lambda^2$, which is an order of magnitude smaller than those of conventional FP SAW resonators with Bragg mirrors. While air damping and thermoelastic damping are the primary limits of the $Q$ factors under atmosphere and at room temperature, other sources of loss could also limit the $Q$ factors of our SAW resonators. The radiation of alternating electromagnetic fields associated with the mechanical strains in strong piezoelectric materials such as lithium niobate could contribute to the total loss.

Beyond SAW resonators on bulk lithium niobate, our design can be extended to thin films of lithium niobate on silica substrates and integrated with optical waveguides. SAW resonators provide an alternative way to manipulate rare earth ions in lithium niobate for quantum information[61]. The high-$Q$ SAW resonators pave the way towards hybrid quantum systems, where phonons could play a crucial role in coupling superconducting qubits, controlling solid-state electronic spins, and performing microwave-to-optical conversion.

**Acknowledgement**

We thank Rebecca Cheng, Lingyan He, Mengjie Yu, and Bernd Kohler for fruitful discussions on manuscript. L.W. is currently at California Institute of Technology. This work was supported by the STC Center for Integrated Quantum Materials, NSF Grant No. DMR-1231319, Army Research Laboratory Center for Distributed Quantum Information Award No. W911NF1520067, and the microwave microscopy was supported by NSF Grant No. DMR-1707372. This work was performed in part at the Center for Nanoscale Systems (CNS), Harvard University. **Author contribution:** L.S. and S.M. fabricated and measured the devices with contributions from other authors. L.Z. and K.L measured the SAW mode profiles. L.S. designed the devices with contribution from L.W. and Y.-I.S.. C.W. and M.Z. built the setup for atmosphere measurements and L.S. built the setup for low temperature measurements. E.P. and L.S. performed FIB milling and SEM imaging. M.N.G. performed the AFM measurements. L.S. and S.M. prepared the manuscript with discussions from all authors. M.L. supervised this project.